\documentclass{osa-article}
\journal{boe}
\articletype{Research Article}

\begin{document}

\title{SNR Enhancement in Brillouin Microspectroscopy using Spectrum Reconstruction}

\author{YuChen Xiang,\authormark{1} Matthew R. Foreman,\authormark{1} and Peter T\"{o}r\"{o}k\authormark{1,2,*}}

\address{\authormark{1}Blackett Laboratory, Department of Physics, Imperial College London, Prince Consort Road, London, SW7 2AZ, UK\\
\authormark{2}School of Physical and Mathematical Sciences (SPMS), Nanyang Technological University, Singapore}

\email{\authormark{*}peter.torok@ntu.edu.sg} %% email address is required

\begin{abstract}
Brillouin {spectroscopy can suffer} from low signal-to-noise ratios (SNRs). Such low SNRs can render common data analysis protocols unreliable, especially for SNRs below $\sim10$. In this work we exploit two denoising algorithms, namely maximum entropy reconstruction (MER) and wavelet analysis (WA), to improve the accuracy and precision in determination of Brillouin shifts and linewidth. Algorithm performance is quantified using Monte-Carlo simulations and benchmarked against the Cram\'er-Rao lower bound. Superior estimation results are demonstrated even at low SNRs ($\geq 1$). Denoising is furthermore applied to experimental Brillouin spectra of distilled water at room temperature, allowing the speed of sound in water to be extracted. Experimental and theoretical values were found to be consistent to within $\pm1\%$ at unity SNR.
\end{abstract}

%%%%%%%%%%%%%%%%%%%%%%%%%%  body  %%%%%%%%%%%%%%%%%%%%%%%%%%
\section{Introduction}
Brillouin spectroscopy is a near century old technique relying on the scattering of incident photons from thermally excited acoustic fluctuations in a medium (i.e. phonons) \cite{Fabelinskii1968}. By virtue of energy conservation, the scattered photons can possess a different frequency to those incident if a phonon is either created or annihilated as part of the interaction. Measurements of this frequency shift, requiring a spectrometer with sufficiently high spectral resolution and throughput, can elicit useful information about phonon velocities and have therefore been extensively employed to study the elastic properties of materials \cite{Lee1986,Speziale2014}. Moreover, study of the linewidth of the  Brillouin spectral peaks enables measurement of phonon lifetimes thereby also enabling viscous characteristics to be quantified \cite{Xu2003,Akilbekova2018}.  More recently, Brillouin spectroscopy has seen a renaissance as it has developed from {a spectroscopic technique into a more powerful hyper-spectral imaging modality} in which mechanical information is mapped with micron level resolution \cite{BIkoski,ScarcelliYun2008,BRI_Scarcelli2015a,BRI_Scarcelli2015b,plaques2015}. This is particularly attractive for its applicability in, for example, \emph{in vivo} diagnostics and cellular imaging (see e.g. \cite{BRI_Meng2016} for a recent review of biomedical applications).   

One drawback of using {spontaneous Brillouin scattering}, however, is the intrinsically low signal-to-noise ratio (SNR) which in turn leads to relatively long image acquisition times. This issue is further exacerbated in fibre-based systems and remains one of the main challenges in designing a clinical Brillouin endoscope device \cite{xiang2017}. {The advent of improved instrumentation, for example, tandem Fabry-P\'erot interferometers, has aimed to address this challenge. Although a spectral contrast of up to $10^{15}$ can be achieved \cite{Sandercock1982}, the acquisition time is nevertheless on the order of seconds per spectrum \cite{Mattana2018} and is hence less suitable for in vivo applications.} Whilst Brillouin measurements can in principle still be made in spite of weak signal strengths, it has been shown that the resulting accuracy and achievable precision of extracted frequency shifts or linewidths are dependent on the SNR \cite{Texas2018,Foreman2019}. Stimulated Brillouin scattering  offers an alternative means to overcome this limitation, and has been used for example in combination with heterodyne detection to efficiently measure large tissue ($\sim 4$~mm) samples \cite{Remer2016a}. Although a promising addition to the Brillouin toolbox allowing large volume measurements to be made in a practical amount of time, the use of spontaneous Brillouin scattering remains the more attractive option for \emph{in vivo} applications because it requires lower incident power, which is in turn more cost-efficient and helps avoid photo-damage to the sample. Consequently, mitigation of low SNR in spontaneous Brillouin spectroscopy is highly desirable. 

Numerous techniques to increase the SNR of a Brillouin system optically can be found in the literature, using for example adaptive optics \cite{Edrei_AO2018}, heterodyning \cite{Ballmann2015} or interferometric filtering \cite{BRI_Carl2016}. Here, however, we present a complementary software based approach, which can be used either independently or in conjunction with existing experimental techniques. By adopting a purely signal processing approach we therefore enable Brillouin measurements to be made in systems that are typically more challenging to optimise optically e.g. an endoscope. 
There exist numerous algorithms for signal reconstruction from noisy data. Here, we restrict our attention to the maximum entropy reconstruction (MER) and wavelet analysis (WA) techniques, due to their applicability in a wide scope of experimental situations, particularly in experiments with low SNR and little knowledge of the sample.  These methods are introduced in Sections~\ref{sec:MER} and \ref{sec:WA} respectively.  The fundamental precision limit given by the Cram\'er-Rao lower bound is introduced in Section ~\ref{sec:FI}, which is then subsequently used to benchmark the performance of the proposed numerical schemes as a function of SNR in Section~\ref{sec:num_results}. Performance improvements relative to more conventional, pure Lorentzian fitting strategies are also presented. Finally, we apply the reconstruction algorithms to experimental Brillouin spectra of distilled water in Section~\ref{sec:exp_results}. {Note that throughout this work we define the SNR as the ratio of the spectral intensity to the standard deviation of additive noise sources, such as detector noise.}

\section{Numerical methods}\label{sec:methods}
Inverse problems are commonly encountered in optics and often require the fitting of empirical models to extract useful information from otherwise incomplete and noisy data   \cite{Anastasio2016}. Within the context of Brillouin spectroscopy, the frequency shifts (or linewidths) of the inelastic spectral peaks are commonly extracted through least-square fitting of multiple peaks (typically Lorentzians) to noisy intensity data taken with a spectrometer. This is similar to the well-known estimation problem encountered in localisation microscopy, for which the position of individual fluorophores is estimated to extract spatial information with super-resolution \cite{SML_Betzig,SML_Hess,SML_Rust}.  Existing estimation techniques can be loosely categorised as fitting or non-fitting, depending on the \emph{a priori} information that is available to the algorithm \cite{SML_Small2014}. The family of algorithms that fit to a known model, such as least-square fitting, usually produce more reliable results at the cost of requiring that some initial parameters are known beforehand. There are also non-fitting methods, such as centroid \cite{SML_Berglund2008} and learning-based \cite{SML_Colabrese2018,Ojha2016} techniques, which require no or minimal information to be known  \emph{a priori}, making them potentially faster and more robust, albeit they are frequently not applicable in all noise regimes \cite{SML_Small2014a}.  Although these are successful strategies in their own right, the resulting solutions are susceptible to noise and especially in the low SNR regime can suffer considerably from bias and poor precision. Noise reduction algorithms can, however, be applied to the collected data before processing so as to enable improved parameter extraction. Linear or nonlinear smoothing filters, for example, are standardised pre-processing procedures that are applied to Raman and infra-red spectral data \cite{SavitzkyGolay1964,Kawata1984} with reasonably good SNR. Although computationally more demanding, more complex algorithms, such as statistical estimators \cite{Bonnet1992,Lasch2012} and principal component analysis \cite{Li2016}, can often reduce noise whilst preserving spectral features to a greater degree. The effectiveness of the techniques described so far either require higher SNRs or certain prior knowledge and assumptions. In this work we therefore consider reconstruction techniques which require minimal prior knowledge and are able to successfully estimate spectral parameters even at low SNRs, namely maximum entropy reconstruction and wavelet analysis. The former is noted for being effective even at low SNRs, whereas the latter is a `tried-and-true' signal denoising method that is especially effective for data degraded by broadband noise. Each method is presented below and may be used independently or in tandem to construct a versatile denoising scheme that can handle a variety of datasets from Brillouin experiments. 

\subsection{Maximum entropy reconstruction (MER)}\label{sec:MER}

Experimental data in spectroscopic applications typically comprises of a series of $N$ discrete intensity values, which we here denote by $d_i$ ($i = 1,\ldots,N$). When spatially or angularly dispersive spectrometers are used, each datum corresponds to the intensity recorded on each of the individual pixels of a segmented detector. For scanning etalon type spectrometers, the data correspond to intensities recorded sequentially in time for each etalon configuration. Observed spectra are however corrupted by noise and thus the determination of the ground truth spectrum is an ill-posed inverse problem. A criterion is therefore required to select a unique spectrum from all possible solutions. The MER method provides one such selection criterion \cite{Scharf1991}. Specifically, the MER technique selects the spectral distribution (described by $f_i$, $i=1,\ldots, N$) which maximises the associated entropy
\begin{align}
S = -\sum_{i=1}^N f_i \log f_i
\end{align}
subject to  the experimental data, i.e. 
\begin{align}
d_i = \sum_{j=1}^N R_{ij} f_j   \label{eq:MERconstraint1}
\end{align}
where $R_{ij}$ is the system response function. For dispersive spectrometers, the entropy can be physically interpreted as the number of bits of information needed to encode on which pixel a single photon was detected \cite{MER_Skilling1984}. 

Practically speaking, in the presence of noise Eq.~\eqref{eq:MERconstraint1} is too restrictive and a solution to the maximisation problem does not exist in general due to the noise degradation of the data. As such it is more appropriate to relax the constraint to allow for an error between the reconstruction and the experimental data. In the presence of additive Gaussian noise for example, the normalised mean square error,
\begin{align}
\chi^2 = \frac{1}{N} \sum_{i,j=1}^N \frac{(R_{ij} f_j - d_j)^2 }{\sigma^2_j},
\end{align}
 is $\chi^2$ distributed \cite{MER_Ables1982, MER_Gull1978}, where $\sigma_j^2$ is the noise variance on pixel $j$.  Maximising the entropy, $S$, subject to the weaker constraint $\chi^2 = \chi_0^2$, where $\chi_0$ is a constant, can therefore allow a solution to be found. Through variation of $\chi_0$, the confidence level of the reconstruction can be varied, with a $95\%$ confidence interval corresponding to $\chi_0^2 \approx 1$ \cite{statsbook}. Imposition of the constraint on $\chi^2$ can be achieved using the method of Lagrange multipliers \cite{Jackson1998} whereby we construct the Lagrangian function
\begin{align}
Q=S-\lambda(\chi^2-\chi_0^2)\label{eq:Q}
\end{align}
where $\lambda$ is a positive Lagrange multiplier. The Lagrangian, $Q$, can be maximised with respect to $\mathbf{f} = [f_1,f_2,\ldots,f_N]$ for any fixed value of $\lambda$. Formally the maximum entropy solution corresponds to the value of $\lambda$ for which $\chi(\mathbf{f})^2 = \chi_0^2$. In practice, however, we have found that similar solutions are found over a broad range of $\lambda$ such that to improve algorithm speed we use a fixed value of $\lambda$ and it is only changed as an initial parameter as the SNR of the data varies. Maximisation of $Q$ can then be performed by iteratively updating $f_i$ until the extremum is reached. A simple update strategy is that based on gradient ascent  \cite{Paper2018} whereby
\begin{align}
\mathbf{f}^{n+1} = \mathbf{f}^n + \mu  \mathbf{P}
\end{align}
where $n$ denotes the iteration number and here $\mathbf{P} = \nabla Q$ is the gradient of $Q$ with respect to ${\mathbf{f}}$. When $\mu$ is chosen appropriately convergence of $\mathbf{f}$ to the value which maximises $Q$ is guaranteed, thus solving the problem:
\begin{equation}
\mbox{max}\big\{ Q(\mathbf{f}) ~|~\mathbf{f}\in \mathbb{R}_+^n\big\}.
\end{equation}
Convergence of this method can, however, be improved by modifying the search direction $\textbf{P}$ using the conjugate gradient technique \cite{Synman2005}. In this method, instead of using $\nabla Q$ to update $\mathbf{f}$, a direction that is conjugate to a set of $r\leq n$ previous search directions  is used. We note, for each iteration $n$, it is possible to find a value of $\mu = \mu_n$ that allows $Q$ to be locally maximised within the subspace spanned by the $r$ previous search directions using the Barzilai-Borwein method \cite{BARZILAI1988}. In this work however we instead use a line search approach with two such `conjugates' (i.e. $r=2$) based on the Wolfe conditions \cite{MER_Wolfe} since this approach is less computationally demanding. Specifically, for each iteration $n$ our algorithm updates the value of $\mathbf{f}$ for varying $\mu_n$ until the Wolfe conditions are met. The process is iterated over $n$ until a termination condition is met, namely that $\frac{1}{2}|\frac{\nabla S}{|\nabla S|}-\frac{\nabla \chi^2}{|\nabla \chi^2|}|^2$ falls below a predefined threshold of $0.01$. This is a direct measure of the balance between the entropy and constraint during the optimisation, and the true maximum entropy solution should yield a value of 0 \cite{MER_Skilling1984}.  Using this algorithm, which we implemented in Matlab, reconstruction of a single Brillouin spectrum takes $\sim500$~ms on a standard office workstation, {albeit run times have not been optimised. The speed and stability of the algorithm can be further improved for example by using more advanced search schemes \cite{MER_Skilling1984}, differentiation algorithms \cite{autodiff2017} or parallelisation \cite{Li1998}}.

Although the MER algorithm can give good results, it is important to note that its ability to produce meaningful spectra is not without bound. The requirement that the reconstructed spectrum is consistent with the data equates to placing an upper limit on the error i.e. $\chi^2\leq \chi_0^2$  as discussed above. If, however, the unconstrained maximum entropy solution for $S$ satisfies this inequality, no meaningful reconstruction is possible since this implies the data is too noisy. This is directly related to the fact that contours of  $\chi^2$ in the vector space defined by $\mathbf{f}$ are convex ellipsoids. Moreover, the entropy surfaces are also convex, implying that if a maximum entropy solution exists it lies on the boundary where $\chi^2=\chi^2_0$ \cite{Smith1985}.

Unlike the majority of localisation methods in microscopy, we note that one advantage of the MER is that it makes no assumptions about the data and in theory requires no \emph{a priori} information. Indeed the algorithm intrinsically assumes we are maximally ignorant about the underlying spectrum \cite{Scharf1991}, whereas localisation algorithms usually function optimally with a well-defined set of assumptions, for example that molecules do not spatially overlap. Nevertheless, it is  possible to include \emph{a priori} information in our reconstruction, if for example the position or width of the likely peaks are approximately known. In turn, such additional information can lead to improved speeds and a higher success rate. Incorporation of such knowledge requires introduction of additional terms into the Lagrangian function.  A careful balance must, however, be struck between consistency of the reconstruction with the actual data and the bias towards the \emph{a priori} information introduced.

\subsection{Reconstruction by wavelet analysis (WA)}\label{sec:WA}
Many conventional noise filtering techniques, such as Fourier filtering or smoothing filters, rely on the fact that the signal of interest and the noise are spectrally diverse. Consequently they can struggle to give good results in the presence of broadband noise. Wavelet analysis (WA) based techniques, however, exploit amplitude filtering in the wavelet domain and thus do not suffer in such scenarios \cite{Mallat2009}. Even with the best detectors, the performance of Brillouin spectroscopy is typically limited by broadband noise \cite{Foreman2019}. In particular, obtained spectra suffer from intensity dependent shot-noise, which for a large enough average signal level can be approximated as following a Gaussian distribution  with a uniform power spectrum \cite{statsbook}.

To illustrate the WA technique consider an additive noise model whereby the observed spectral data can be written in the form
\begin{align}
d_i = \sum_{j=1}^N R_{ij}f_j + n_i
\end{align}
where $d_i$, $R_{ij}$ and $f_j$ are defined as above and $n_i$ denotes a random noise contribution to the $i$th datum ($i= 1,2,\ldots,N$).
This experimental spectrum is first decomposed into a chosen basis of $M$ wavelets using the discrete wavelet transform (DWT) \cite{Griffel1995}, such that
\begin{align}
\widetilde{\mathbf{d}} = \mbox{DWT}[\mathbf{d}] = \widetilde{\mathbf{g}} + \widetilde{\mathbf{n}}
\end{align}
where we have defined $g_i = \sum_{j=1}^N R_{ij}f_j$, $\mathbf{d} = [d_1,d_2,\ldots,d_N]$ and $\widetilde{\mathbf{d}} = [\widetilde{{d}}_1,\widetilde{{d}}_1,\ldots,\widetilde{{d}}_M]$  is the  vector formed from the amplitude coefficients of each individual wavelet (known as a level) in the observed data (similarly $\widetilde{\mathbf{g}}$ and $\widetilde{\mathbf{n}}$ contain wavelet coefficients of each level for the observed spectrum and the noise). Under the assumption that the noise is smaller in amplitude than the spectrum of interest, the noise contribution can be reduced by applying a nonlinear thresholding (commonly referred to as shrinking) of the coefficients of the DWT, each according to a threshold $Q$. Nonlinear shrinkage of the coefficients is such that smaller amplitude coefficients (i.e. those describing the noise) are reduced to a greater extent than those of the spectrum. Thresholding can be performed using either hard or soft limits, although  soft limits are usually preferable in spectroscopy in order to preserve spectral features. By limiting the consideration to only white noise, the thresholding strategy is further simplified as it can be proven statistically that a universal threshold (i.e. the same for each level) can be set \cite{Donoho1995a}. Finally, an inverse DWT is performed on the `denoised' wavelets, yielding a reconstructed spectrum \cite{wavelets_Srivastava2016}. Although WA is a quick and easy means of extracting spectral features from noisy data, at lower SNRs the argument for universal thresholding is somewhat questionable and the process is then seen to fundamentally rely on some \emph{a priori} knowledge of the shape and location of anticipated features thus rendering the process inevitably subjective. In these cases, there is also a limitation which comes from the assumption that the signal is strong enough compared to the noise level which allows the separation in amplitude, meaning that useful denoising may not always be possible. 

In this work we employ the \emph{Matlab Wavelet Analyzer} application and use an adaptive thresholding algorithm to denoise spectra based on the SNR which allows reconstruction of a single spectrum in less than 10~ms. The threshold is generally set to be level-independent and follows the relationship given in Ref. \cite{Donoho1995a}:
\begin{equation}
    \text{Universal~threshold~}Q=n\sqrt{2\ln(N)/N}
\end{equation}
where $n$ is the experimentally determined noise level, which is constant in the case of white noise. In principle, even in the presence of intensity-dependent noise i.e. shot noise, this can be set to adaptively vary for different levels. For example, it is possible to incorporate level-dependent thresholding schemes where $n$ for each level can be estimated statistically \cite{Donoho1995b}.

\section{Assessing algorithm performance}

\subsection{Informational limit of Brillouin systems}\label{sec:FI}
To assess the performance of denoising algorithms in Brillouin spectroscopy it is necessary to define suitable performance metrics and to have a benchmark to which they can be compared. Given the objective of denoising algorithms, a natural choice is to consider the average SNR in the reconstructed spectra as compared to the unprocessed data. Ultimately, however, Brillouin spectroscopy aims to quantitatively study the scattering induced frequency shifts. So as to better reflect the aims of Brillouin spectroscopy, we hence choose to consider the accuracy and precision to which the frequency shift $\Omega$ can be determined. From a statistical perspective, the accuracy of an estimate of $\Omega$ can be parameterised using the bias, $b_\Omega$, which is defined as the difference between the true value of $\Omega$ as compared to  the average frequency shift $\langle\hat{\Omega}\rangle$ returned from an estimation protocol i.e. $b_\Omega = \langle\hat{\Omega}\rangle - \Omega$. Note here we use the standard statistical notation whereby $\hat{\Omega}$ denotes an estimate of $\Omega$ and $\langle \cdots \rangle$ denotes an ensemble average. The precision can similarly be quantified using the estimator variance $\sigma_\Omega^2 = \langle\hat{\Omega}^2\rangle - \langle\hat{\Omega}\rangle ^2$. Although extensive research quantifying precision limits in localisation microscopy exists \cite{Ober2004,Smith2010} it is worth noting that in the context of inelastic spectroscopy at least two peaks are always needed to estimate the relative frequency shift in contrast to the single peak used in localisation microscopy. Consequently the estimation problem is more akin to determining the separation of two molecules \cite{Ram2006,Tsang2016}. Derivation of the relevant precision limits in the context of Brillouin spectroscopy is given explicitly in Ref.~\cite{Foreman2019}.

 Fundamentally there is no limit as to how small the bias of an estimator can be, however, the limit of precision to which $\Omega$ can be determined follows from the well known Cram\'er-Rao lower bound (CRLB). Specifically, the CRLB bounds the variance, $\sigma^2_\Omega$, of any estimate of the frequency shift according to $\sigma^2_\Omega \geq J^{-1}$, where $J$ is the associated (noise dependent) Fisher information \cite{Scharf1991}. Assuming a dispersion limited spectrometer with a pixelated detector and restricting to Gaussian white noise of variance $\sigma^2$ (as used in our simulations below), the obtainable precision is bounded according to \cite{Foreman2019}
\begin{align}
{\sigma^2_\Omega} \geq  \frac{\pi \Delta}{4X^2 } \frac{(\alpha \Gamma_\pm + \gamma)^3 }{\text{SNR}^2} \frac{(1+2I_\pm)^2}{\alpha^2 I_\pm^2}. \label{eq:CRLB}
\end{align}
Here $\alpha$ is the linear scale factor between the spatial position $x$ on the detector and optical frequency $\omega$ (i.e. $x \sim \alpha \omega$), $I_\pm$ is the  intensity of the Stokes and anti-Stokes peaks (assumed the same) relative to the Rayleigh intensity, $\Delta$ is the size of the pixels, $X$ is the width of the detector, $\gamma$ is the FWHM of the spectrometer response function,  $\Gamma_\pm$ is the FWHM of the Brillouin peaks and the average signal to noise ratio on each pixel is $\text{SNR} =  I_{\infty} \Delta / ( X \sigma) $, where $I_{\infty}$ is the integrated spectral intensity and $\sigma^2$ is the noise variance. 

\subsection{Reconstruction of simulated data}\label{sec:num_results}

\begin{figure}[t!]
	\centering
	\includegraphics[width=0.7\columnwidth]{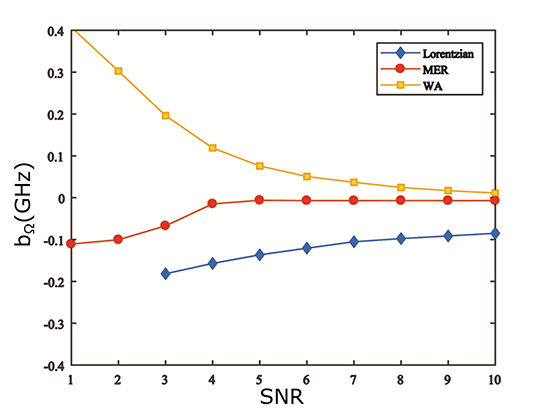}
	\caption{Bias of estimates of the Brillouin shift found by Lorentzian fitting of {simulated} noisy spectral data subject to different denoising algorithms, as calculated using 5000 realisations of simulated noise.
    \label{fig:bias}}
\end{figure}

\begin{figure}[t!]
	\centering
	\includegraphics[width=0.7\columnwidth]{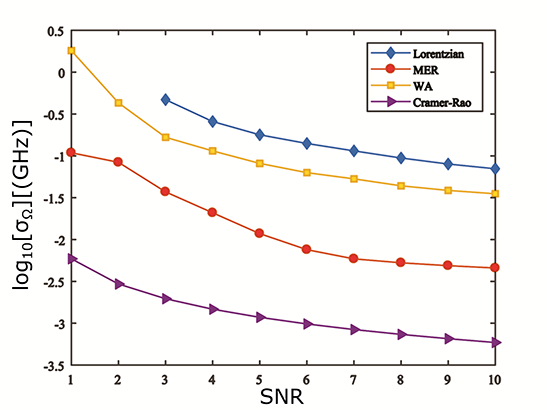}
	\caption{Logarithm of the standard deviation of estimates of the Brillouin shift, found by Lorentzian fitting of {simulated} noisy spectral data subject to different denoising algorithms, as calculated using 5000 realisations of simulated noise. 
		\label{fig:std}}
\end{figure}

With a suitable benchmark in hand, it is now possible to quantitatively assess the performance of the proposed denoising algorithms, by means of Monte Carlo simulations. The ground truth Brillouin spectrum used assumed the Rayleigh (elastic) peak of frequency $\sim534$~THz ($561$~nm) was centred on a detector of total width $16.6$~mm and pixel size $6.5$~$\mu$m, where $120$ pixels were used to correspond to an effective bandwidth of $\sim60$~GHz. Note that these values were chosen to resemble a realistic experimental set-up. In reality, such coarse pixelation of the spectrum may have an adverse effect on the estimation precision {and bias} of the relative shift.  {In particular, pixelation effects become significant when recorded spectral peaks have widths on the order of the pixel size as discussed further in Ref.~\cite{Foreman2019}. To enable a fair comparison between different algorithms we use a fixed pixel size for all simulations.} The spectral width of both the Rayleigh peak and Brillouin peaks was assumed to be $1$~GHz. The Brillouin peaks were taken to have a relative shift of $\pm 10$~GHz, whilst the peak intensity of the (anti)-Stokes and the Rayleigh peak was assumed to be $10^3$ and $10^4$ respectively. For simplicity an ideal system response function was also assumed (i.e. $R_{ij} = \delta_{ij}$ where $\delta_{ij}$ is the Kronecker delta, whereby $\gamma = 0$). A total of 5000 independent realisations of white Gaussian noise (of varying $\mbox{SNR} \leq 10$) were added to the `true' spectrum. The spectral parameters (most importantly ${\Omega}$) of each noisy spectrum were then estimated using Lorentzian fitting before and after the denoising algorithms were applied. The bias and standard deviation of the estimated Brillouin shift were then calculated and are shown in Figs.~\ref{fig:bias} and \ref{fig:std} respectively as a function of SNR. Note that due to poor performance at lower SNRs we found it necessary to incorporate prior information of the number and positions of spectral peaks into the MER algorithm i.e. the approximate position and width of the Lorentzian function, as discussed further in Ref.~\cite{MER+RAM_Craggs1996}. 

Inspection of Figs.~\ref{fig:bias} and \ref{fig:std} demonstrates that both denoising algorithms generally facilitate more accurate and precise estimation of the Brillouin shift. Furthermore, we note that the MER produces superior performance to the WA based method, especially at low SNRs. Although performance of all algorithms are found to degrade at lower SNRs, even for an SNR of unity, the MER  allows $\Omega$ to be determined with a bias of approximately 1\% and an uncertainty of $\sim 1$\%.  It should however be noted that spectrum reconstruction was not possible for all realisations in this case as circa 20\% of the generated spectra did not possess ME solutions and were regenerated until the total number of realisations were fulfilled. In contrast, the unconstrained Lorentzian fitting algorithm without denoising was unable to retrieve any meaningful information of the spectra at the same SNR. Both the bias and the uncertainty were well outside of reasonable bounds for SNR $\le2$, thus they were not plotted for comparison. Fitting performance of unprocessed spectra at an SNR of 10 is still notably degraded by noise yielding a bias (standard deviation) of $\sim 0.9$\% ($\sim 0.7$\%) in estimates of the Brillouin shift.
Similar trends were also found (data not shown) when estimating the linewidths of the obtained Brillouin peaks. In this case, MER also yielded the smallest bias relative to the ground truth width (1 GHz) and the best precision, whilst WA again outperforms Lorentzian fitting. As an indicative example, for an $\mbox{SNR}=5$, pure Lorentzian fitting performed poorly giving an average linewidth of $7.29\pm1.07$ GHz, whilst both the MER and WA were noticeably superior, producing linewidths of $1.08\pm0.04$ and $2.40\pm0.13$ GHz respectively. 

With respect to the WA method, we note that whilst its utility for denoising is evident from Figs.~\ref{fig:bias} and \ref{fig:std} for SNRs $\gtrsim 3$, at lower SNRs it is unable to function reliably. This degradation in performance is, however, unsurprising because in this high noise regime the strength of the noise is comparable to the signal in contradiction to the assumptions of the WA denoising algorithm. As such, the shape of the reconstructed spectrum becomes very sensitive to the precise manner in which the coefficients threshold is chosen and thus erroneous shifts and distortions of the individual peaks can result.  This is particularly detrimental to accurate retrieval of the Brillouin linewidth. Wavelet shrinkage can hence degrade potentially useful information from the signal. Finally we note that although both considered denoising algorithms produce better estimates of the Brillouin shift and linewidth, there remains further scope for improvement since neither achieve the CRLB, although the difference between the theoretical lower bound and the simulation results may be partially accounted for by pixelation effects mentioned above. Statistical methods such as the method of Maximum Likelihood Estimation (MLE) \cite{MER_Vella2018} can routinely achieve the CRLB but have yet to be applied to Brillouin spectroscopy. This remains an interesting area for future study.

\section{Denoising of experimental spectra}\label{sec:exp_results}

\begin{figure}[t!]
	\centering
	\includegraphics[width=0.7\columnwidth]{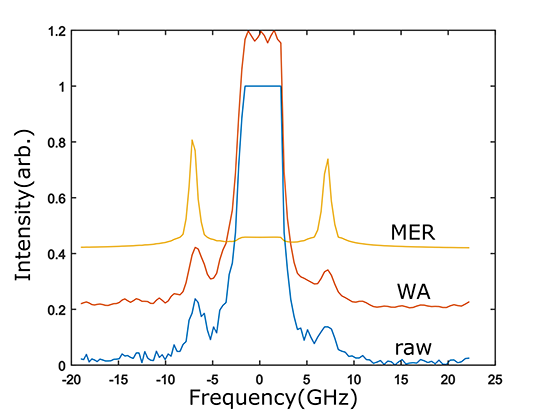}
	\caption{Example of a typical unprocessed {experimental} Brillouin spectrum of distilled water obtained using a 100~ms acquisition time ($\mbox{SNR}\sim 5$) (blue). Corresponding reconstructed spectra as found using the WA (orange) and MER (yellow) algorithms are also shown. Note that spectra have been vertically shifted to improve visibility.  \label{fig:spectra}}
\end{figure}

To demonstrate the applicability of the denoising algorithms to real spectral data, Brillouin spectra of pure water at room temperature were taken using a custom-built Brillouin set-up (see \cite{Karampatzakis2017} for full details). Variation of the SNR was realised through use of different exposure times. A weak incident power of $\sim 5$~mW was also used to operate in the low SNR regime. Typical examples of the raw and denoised spectra for an SNR of $\sim 5$ are shown in Fig.~\ref{fig:spectra}. {Critically, the central Rayleigh peak seen in Fig.~\ref{fig:spectra} is saturated which is a feature commonly encountered in Brillouin spectroscopy. This meant that although MER was sometimes possible, even at SNRs as low as 1, the resulting reconstruction did not generally lie within a $95\%$ confidence error due to the large deviation in peak shape that saturation produces. To help mitigate the issues associated with the saturated Rayleigh peak, interferometric suppression of the central Rayleigh peak \cite{BRI_Carl2015} was used, which in turn improved the contrast of the Brillouin peaks, algorithm convergence rates and reliability of the fitting results. Importantly, Rayleigh suppression can be fully accounted for in MER through the form of the \emph{a priori} information introduced. If the exact form of the suppressed Rayleigh peak is not known \emph{a priori}, the Rayleigh peak can alternatively be excluded from the MER entirely, as is the case for the data shown in Fig.~\ref{fig:spectra}.}

Upon calibrating the spectrometer and fitting of the (denoised) experimental spectra, the Brillouin frequency shift was found and the corresponding speed of sound determined \cite{Prevedel2019}. Table~\ref{tab:sound} compares the values obtained using each denoising procedure for different SNRs (or equivalently exposure times). The respective RMS fitting error in percentage is also listed to indicate the relative fitting accuracy. {Note that poorer SNRs required stronger interferometric suppresion of the Rayleigh peak. Consequently, algorithm performance for different SNRs is not directly comparable, however, similar relative improvements between algorithms for a given SNR are evident.}

\begin{table}[b!]
	\begin{center}
		{
			\begin{tabular}{|l||c|c|c|}
				\hline
				& $\mbox{SNR}\sim 5$ & $\mbox{SNR}\sim2$  & $\mbox{SNR}\sim 1$ \\
				&(100~ms) &  (50 ms) & (30 ms) \\ \hline\hline
				MER       & $1485.6(2.4\%)$    & $1490.9(2\%)$   & $1493.2(3\%)$   \\ \hline
				WA        & $1481.0(4\%)$    & $1450.2(5\%)$   & $1508.0(13\%)$  \\ \hline
				None & $1494.5(34\%)$  & $1488.8(27\%)$  & $1612.7(37\%)$ \\ \hline
			\end{tabular}
			\caption{Speed of sound in distilled water as obtained from Lorentzian fitting of experimental Brillouin spectra subject to the MER and WA algorithms as compared to no pre-processing. \label{tab:sound} }
		}
	\end{center}
\end{table}

  Whilst the speed of sound is sensitive to a range of parameters, it is expected that the speed of sound in distilled water at the laboratory temperature of $22^{\circ}C$ is around $1490$~m/s \cite{BRI_water}. The MER denoising algorithm was thus once again found to perform best out of the methods considered here, producing experimental values of the speed of sound with low uncertainties  and in good agreement with the expected values. Whilst WA also gave good agreement at higher SNRs, a significant bias is seen to occur at low SNRs in agreement with the findings of Fig.~\ref{fig:bias}. Finally, direct Lorentzian fitting was found to produce large errors for all SNRs considered {due to the poor quality of the spectrum}, such that no meaningful conclusions can be made.

\section{Conclusion}
In this work, we have proposed use of two denoising algorithms to improve the accuracy and precision of data analysis in Brillouin spectroscopy. As far as we are aware, this is the first time that such reconstruction schemes have been used to enhance Brillouin spectroscopic data. Specifically, we have discussed both a MER and WA based data processing chain. Through Monte Carlo simulations we have demonstrated that both algorithms enable reduced bias and greater precision when estimating both the frequency shift induced by Brillouin scattering and the associated phonon lifetime. For example, at least a three-fold precision gain was demonstrated when estimating the Brillouin shift when WA was used, whereas up-to an order of magnitude improvement was demonstrated using MER. Indeed, MER was generally found to outperform WA, however, both denoising algorithms enabled superior performance down to SNRs of $\gtrsim 1$, a regime in which conventional line fitting fails. {Although our simulations assumed white Gaussian noise, as is common place in many experimental scenarios, more complex noise regimes such as coloured and structured noise can also be accounted for through introduction of additional \emph{a priori} information, use of a pre-whitening processing step or a more elaborate thresholding strategy.}

 Experimental Brillouin spectra of distilled water were also used to verify the denoising algorithms in a real-world scenario. In particular, improved data analysis was demonstrated with the experimentally determined speed of sound agreeing with theoretical values to within $\sim 1\%$. In  conclusion,  we  have  thus shown  that  a quantifiable SNR  enhancement  of  noisy  Brillouin  data  is  easily achievable through  numerical  methods,  thereby  unlocking  the  potential for faster and more reliable measurements. Critically, by adopting a software based approach such improvements are complementary to any technical experimental gains and thus represent an attractive option for Brillouin applications where optical design is impractical, for example in the case of Brillouin endoscopy, {or as a means to improve the acquisition speed by enabling operation at lower SNRs}. Further development of these denoising schemes can lead to yet further gains, achieving the CRLB, and thus promise to become a useful tool in Brillouin spectroscopy and an embedded part of the data analysis routine in this field.

\section*{Funding}
UK Engineering and Physical Sciences Research Council; The Royal Society (UK)

\section*{Disclosures}
\noindent The authors declare that there are no conflicts of interest.

%%%%%%%%%%%%%%%%%%%%%%% References %%%%%%%%%%%%%%%%%%%%%%%%%

\end{document}